\documentclass[
               twocolumn,
               noshowpacs,          
               nopreprintnumbers,     
               aps,                 
               prl,                 
               letter,             
	       superscriptaddress,  
               nofootinbib,         
               tightenlines,        
               floats,floatfix    
               ]{revtex4-1}
\usepackage{amsmath, amsfonts, amsthm, amssymb, graphicx}
\usepackage{color}
\usepackage[utf8]{inputenc}
\usepackage{lipsum}
\usepackage{centernot}
\usepackage{ulem}
\usepackage{feynmp}

\def\be{\begin{equation}}
\def\ee{\end{equation}}
\def\ba{\begin{eqnarray}}
\def\ea{\end{eqnarray}}
\def\beastar{\begin{eqnarray*}}
\def\eeastar{\end{eqnarray*}}       

\def\del{\partial}
\def\bdm{\begin{displaymath}}
\def\edm{\end{displaymath}}

\def\bq{\begin{quote}}
\def\eq{\end{quote}}

 at 10truept
\newcommand{\mut}{\ensuremath{\tilde{\mu}}}                                                                                                                            
                                                                                                                            
\newcommand{\phib}{\ensuremath{\bar{\phi}}}                                                                                                                            
\newcommand{\grad}{\ensuremath{\vec{\nabla}}}

\newcommand{\beq}{\begin{equation}}
\newcommand{\eeq}{\end{equation}}
\newcommand{\bea}{\begin{eqnarray}}
\newcommand{\eea}{\end{eqnarray}}
\newcommand{\beqa}{\begin{eqnarray}}
\newcommand{\eeqa}{\end{eqnarray}}

\def\L{{\cal L}}
\def\E{{\cal E}}
\def\K{{\cal K}}

\newcommand{\citeGW }{\cite{Spanish,Sak,vern,pedro}}

\def\ltap{\ \raise.3ex\hbox{$<$\kern-.75em\lower1ex\hbox{$\sim$}}\ }
\def\gtap{\ \raise.3ex\hbox{$>$\kern-.75em\lower1ex\hbox{$\sim$}}\ }
\def\gl{\ \raise.5ex\hbox{$>$}\kern-.8em\lower.5ex\hbox{$<$}\ }
\def\roughly#1{\raise.3ex\hbox{$#1$\kern-.75em\lower1ex\hbox{$\sim$}}}

\begin{document}

\title{Dark energy after GW170817, revisited}

\author{Edmund J. Copeland} 
\email{ed.copeland@nottingham.ac.uk}
\affiliation{School of Physics and Astronomy, 
University of Nottingham, Nottingham NG7 2RD, UK} 
 \author{Michael Kopp}
 \email{kopp@fzu.cz}
 \affiliation{CEICO, Fyzikální ústav Akademie v\v{e}d \v{C}R, Na Slovance 2, 182 21 Praha 8, Czechia}
\author{Antonio Padilla} 
\email{antonio.padilla@nottingham.ac.uk}
\affiliation{School of Physics and Astronomy, 
University of Nottingham, Nottingham NG7 2RD, UK} 
\author{Paul M. Saffin}
\email{paul.saffin@nottingham.ac.uk}
\affiliation{School of Physics and Astronomy, 
University of Nottingham, Nottingham NG7 2RD, UK} 
 \author{Constantinos Skordis}
 \email{skordis@fzu.cz}
 \affiliation{CEICO, Fyzikální ústav Akademie v\v{e}d \v{C}R, Na Slovance 2, 182 21 Praha 8, Czechia}

\date{\today}

\begin{abstract}
We revisit the status of scalar-tensor theories  with applications to dark energy in the aftermath of the gravitational wave signal GW170817  and its optical counterpart GRB170817A.  At the level of the cosmological background, we identify a class of theories, previously declared unviable in this context, whose anomalous gravitational wave speed is proportional to the scalar equation of motion.  As long as the scalar field is assumed not to couple directly to matter, this raises the possibility of  compatibility with the gravitational wave data,  for  any cosmological sources, thanks to the scalar dynamics.   This newly ``rescued" class of theories includes examples of  generalised quintic galileons from Horndeski theories.  Despite the promise of this leading order result, we show that the loophole  ultimately fails when we include  the effect of large scale inhomogeneities. 
  \end{abstract}
\maketitle

\section{Introduction}
The observation of the neutron star merger GW170817 at redshift $z \sim 0.01$ and its optical counterpart GRB170817A  \cite{gravitational wave , GRB1,GRB2,GRB3,GRB4} has provided spectacularly strong constraints on the relevance of modified gravity  \cite{myreview,penn, lav} to understanding the dynamics of the late Universe \cite{dark}.  Generic interactions between the massless spin 2 and additional light fields can cause the gravitational wave to propagate  through the cosmological background at speeds different from its electromagnetic counterpart, even when it passes through overdense regions where so-called screening mechanisms might be expected to operate \cite{Jimenez,lucas1, lucas2}.  In the light of the LIGO/Virgo observation, a careful analysis of this effect in a wide class of scalar tensor theories has led to a dramatic reduction in the landscape of modified gravity models that are relevant for dark energy and remain observationally viable at late times \citeGW  (see also \cite{Amendola, Langlois, Marco,boss, Kase}). 

To avoid being drawn into an erroneous narrative as to the fate of modified gravity, it is important to properly state the implications of    \citeGW. Although a significant number of scalar tensor interactions were ruled out, they did not rule out everything, even in the context of dark energy. For example, conformal couplings to curvature,  as in Brans Dicke \cite{bd} or chameleon models  \cite{cham1, cham2},  remain viable, as do  so-called Kinetic Gravity Braiding models \cite{kgb}. Furthermore, there is nothing to stop us from including the forbidden interactions  as long as they are suppressed by some heavy scale that renders them irrelevant in the late Universe.  Of course, this latter consideration weakens the motivation for considering such interactions in the first place.  

Going beyond these clarifications, there are  also reasons to revisit the conclusions  of \citeGW . In particular, it was recently noted that the frequency  scales of the neutron star event lie close to the strong coupling scale associated with many dark energy models. If they are known, ultra violet effects could impact the speed of the gravitational wave, and if they are not known, one is attempting to constrain a theory  outside of its regime of validity \cite{scott}. Of course, such a manifest breakdown of the low energy description could also be relevant/problematic to theories that rely on the Vainhstein effect  to pass solar system tests  \cite{unit,ippo}  (for further understanding of Vainshtein screening, see  \cite{vain1,vain2,vain3}; for discussions on their UV completions see \cite{amps, tolley, oranges}). In this paper, we consider the possibility of a different loophole: that the speed of the gravitational wave is set to unity {\it dynamically}\footnote{See \cite{betoni} for related ideas for which the speed of the gravitational wave approaches unity dynamically due to a non-minimal coupling between dark energy and dark matter.}. More precisely, in the context of scalar tensor theories, we identify new scenarios in which the deviation from unity is proportional to the scalar field equation of motion on the cosmological background. As long as the scalar is decoupled from the matter sector directly, its field equation will always vanish identically ensuring exact agreement with the LIGO/Virgo observations at this order.  However, as we will see, there is no extension of  this result  including  the effect of large scale  inhomogeneites.

Within the Horndeski class \cite{horn,george},  so-called  ``$\L_5$" operators, cubic in second derivatives of the scalar, were previously thought to be excluded by the gravitational wave data already at the level of the background cosmology.  However, we shall present an explicit example of a theory in which such an operator is present and yet the speed of the gravitational wave is unity at the background level thanks to the vanishing of the  scalar equation of motion.  In going beyond Horndeski \cite{bh1,bh2}, we find other interactions that can be rescued from the forbidden zone along the same lines.  However, it is only the Horndeski example that survives additional constraints coming from the decay of the wave into dark energy fluctuations \cite{decay}. As stated above, to rule out this newly rescued Horndeski class, we need to consider the effects of large scale inhomogeneities.

\section{LIGO/Virgo revisited}
The sight and sound  of the neutron star merger detected on August 17, 2017 has constrained the speed of gravitational and electromagnetic waves through the cosmological medium at late times,  to be identical to an accuracy of  the order $10^{-15}$.  Although these speeds are indeed identical in General Relativity, this is not the case in generic modified gravity models where the additional fields can possess a non-trivial cosmological profile that pushes the tensor mode off the light cone. To illustrate this,  we consider a wide class of scalar tensor theories, including Horndeski \cite{horn,george} and beyond Horndeski  \cite{bh1,bh2} interactions, given by the action $S=\int d^4 x  \sqrt{-g}\sum_{n=2}^5 {\cal L}_n$, 
where 
\small
\ba
&&\L_2 = G_2(\phi, X)   \nonumber \\
&&\L_3= G_3(\phi, X) \Box \phi  \nonumber \\
&&\L_4 = G_4(\phi, X)R-2G_{4, X} \nabla_{[\mu_1}\nabla^{\mu_1} \phi \nabla_{\mu_2]}\nabla^{\mu_2} \phi  \nonumber \\
&&  \qquad +F_4(\phi, X)\epsilon^{\mu\nu\rho}{}_\sigma \epsilon^{\mu'\nu'\rho'\sigma}\nabla_\mu \phi \nabla_{\mu'} \phi \nabla_{\nu}\nabla_{\nu'} \phi \nabla_\rho \nabla_{\rho'} \phi  \nonumber \\
&& \L_5 = G_5(\phi, X) G_{\mu\nu} \nabla^\mu \nabla^\nu \phi
 +\frac{G_{5,X}}{3} \nabla_{[\mu_1}\nabla^{\mu_1} \phi \nabla_{\mu_2}\nabla^{\mu_2} \phi  \nabla_{\mu_3]} \nabla^{\mu_3} \phi   \qquad \nonumber \\
&&\qquad  + F_5(\phi, X)\epsilon^{\mu\nu\rho\sigma}\epsilon^{\mu'\nu'\rho'\sigma'}\nabla_\mu \phi \nabla_{\mu'} \phi \nabla_{\nu}\nabla_{\nu'} \phi \nabla_\rho \nabla_{\rho'} \phi \nabla_{\sigma}\nabla_{\sigma'} \phi \qquad  \nonumber
\ea
\normalsize
Here we have a metric $g_{\mu\nu}$ with corresponding covariant derivative $\nabla_\mu$, Ricci scalar $R$ and Einstein tensor $G_{\mu\nu}$. We have  a scalar field $\phi $ and define its canonical kinetic operator  $X=g^{\mu\nu }\del_\mu \phi \del _\nu \phi$.  The symbol  $\epsilon^{\mu\nu\rho\sigma}$ is the totally antisymmetric Levi-Civita tensor while the  square brackets denote antisymmetric combinations defined without the usual  factors of $1/n!$ $G_{4, X}$ and $G_{5, X}$ denote derivatives of the potentials with respect to $X$. Despite the higher order nature of these theories one can avoid propagating additional degrees of freedom associated with Ostrogradski ghosts \cite{ostro}. In particular, the theory will propagate one scalar and two graviton degrees of freedom in each of the following cases \cite{dhost3}: the Horndeski class \cite{horn} with second order field equations for which $F_4=F_5=0$; beyond Horndeski \cite{bh1,bh2} with $\L_4=0, G_{5, X} \neq 0$ {\it or} $\L_5=0, G_4-2XG_{4, X} \neq 0$; beyond Horndeski  with both  $\L_4 \neq 0, \L_5 \neq 0$ and a degeneracy condition $XG_{5,X}F_4 =3F_5\left[ G_4-2XG_{4,X} -(X/2)G_{5,\phi}\right]$. In the latter case $F_4$ and $F_5$ are generated by the same disformal transformation \cite{dhost3, bh1,bh2}.

To proceed, we take a spatially flat cosmology, $ds^2=-dt^2+a(t)^2 d\bf x^2$, with a homogeneous scalar, such that we also have  $X=-\dot \phi^2$. We shall further  assume that matter is minimally coupled to the metric  with no direct coupling to the scalar, so that the scalar field equation has no external source. The form of the corresponding field equations can be obtained using \cite{bh1}, by varying the minisuperspace Lagrangian with respect to the lapse  function and the scalar field, then setting the lapse to unity. The key point in what follows  is that the scalar equation takes the form ${\cal E}_\phi\equiv \frac{1}{a^3} \frac{\delta  S}{\delta \phi}=A\ddot \phi-B$ where $A=\sum_{i, j} A_{ij}(\phi, X)H^i \dot H^j, B=\sum_{i, j} B _{ij}(\phi, X)H^i \dot H^j$, and $H=\dot a/a$,  with explicit formulae for the $A_{ij}, ~B_{ij}$ given in the appendix. 
 As long as $A$ is non-vanishing this can be used to identify $\ddot \phi=B/A$. In other words, imposing the vanishing of the scalar equation of motion means that we should not treat $\ddot \phi$ as independent of $\phi, X, H, \dot H$.  

The tensor fluctuations on this background are described by the following quadratic action \cite{kob,bh1}
\be
S_T^{(2)}=\frac18 \int dt d^3 x a^3 \left[{\cal G}_T \dot h_{ij}^2-\frac{{\cal F}_T}{a^2} (\vec \nabla h_{ij})^2 \right]
\ee
where 
\ba
&&{\cal F}_T= 2\,G_{{4}} +XG_{5, \phi} -2X \ddot \phi G_{5, X}
\\
&& {\cal G}_T=2G_4-4XG_{4, X}-XG_{5, \phi}+2X^2 F_4 \nonumber \\
&& \qquad  \qquad -2HX\dot \phi(G_{5, X}+3XF_5)
\ea
The speed of the gravitational wave through the cosmic medium is now given by $c_T^2=\frac{{\cal F}_T}{ {\cal G}_T}$, and its deviation from unity by $\alpha_T=c_T^2-1=\frac{{\cal F}_T-{\cal G}_T}{ {\cal G}_T}$.  This is the quantity that, at late times,  is constrained to vanish to order $10^{-15}$ thanks to the neutron star merger. We will therefore require it to be zero. 

In \citeGW, the authors require $\alpha_T$ to vanish on {\it any} cosmological background.  As is elegantly explained  in \cite{vern,pedro} we can write ${\cal G}_T\alpha_T=-2X \ddot\phi G_{5, X}+C(\phi, X)$, where\footnote{Note that we have traded $\dot \phi=\sqrt{-X}$.  We could have chosen the root with opposite sign but this would not affect our conclusions since the subsequent analysis is invariant under $t \to -t$ \label{signchoice}}  
\begin{multline}
C(\phi, X)=4XG_{4, X}+2XG_{5, \phi}-2X^2 F_4 \\
+2HX\sqrt{-X}(G_{5, X}+3XF_5). \nonumber
\end{multline}
  Requiring $\alpha_T$ to vanish for any cosmological background, they demand that it vanishes for any choice of $\ddot \phi$. This imposes {\it two} independent conditions $G_{5, X}=0$ and $C(\phi, X)=0$. Regarding them as partial differential equations for  the (beyond) Horndeski potentials one is then able to greatly constrain the space of scalar tensor theories that are compatible with the gravitational wave data. However, based on our earlier discussion regarding the form of the scalar equation of motion, we  will see that this approach is {\it too} constraining and that  we are  free to use $\E_\phi=0$ to eliminate $\ddot \phi$ in our expression for $\alpha_T$.  Proceeding in this way, we obtain  ${\cal G}_T\alpha_T=-2X\frac{B}{A} G_{5, X}+C(\phi, X)$, then  require $A {\cal G}_T\alpha_T$ to vanish, giving a complicated equation of the form 
  \be
  \sum_{i, j} C_{ij}(\phi, X)H^i \dot H^j=0, \label{cond}
\ee
where $C_{ij}=-2XB_{ij}G_{5, X}+C A_{ij}$. We now demand that \eqref{cond} holds for any choice of energy density, $\rho$,  and pressure, $p$, or in other words, it should hold for all values of $H$ and $\dot H$.  This results in a number of constraints $C_{ij}=0$ that can be treated as a simultaneous set of partial differential equations for the (beyond) Horndeski potentials. To solve them we first note that $A_{i1}$ vanishes for all $i$, in contrast to $B_{i1}$ (see appendix). Imposing $C_{i1}=0$ is therefore equivalent to $G_{5, X} B_{i1}=0$.  This is the fork in the road. On the one hand we can solve this by setting $G_5=G_5(\phi)$. It then follows that $C_{ij}=CA_{ij}$ and since $A$ must be non-vanishing we have that $C=0$. This reduces to the scenario already considered in \citeGW.  Alternatively, however, we may assume that $G_{5, X} \neq 0$, in which case we must have that $B_{i1}=0$.  This leads to a new class of solutions that are compatible with the gravitational wave data, where the (beyond) Horndeski potentials are given as 
 \bea
&& G_2=\frac12 XH_{1, \phi\phi\phi}-XH_{2, \phi\phi}+\K+Xh'(\phi) \qquad \label{G2}\\
&& G_3=XH_{1,\phi\phi X}-2XH_{2, \phi X}+\frac12 H_{1,\phi\phi}-H_{2, \phi}+h(\phi)\qquad  \\
&& G_4=\kappa_G-\frac12 X H_{1,\phi X}+XH_{2, X} \\
&& G_5=H_{1, X}-6\frac{\mu}{\sqrt{-X}} \\
&& F_4=-H_{1,\phi X X}+2H_{2, XX}+\frac{H_{2,X}}{X} \\
&& F_5=-\frac{H_{1, XX}}{3X} \label{F5}
\eea
where $\kappa_G$ and $\mu$ are constants.  $\K(\phi, X)$ is a function of $\phi,  X$,  arbitrary up to the condition $\K_{, X} \neq 0$.  We  use it to  obtain $H_1(\phi, X)$ and $H_2(\phi, X)$ via  the following differential equations,
\bea
&&H_{1, XX} =-2\mu \left[ \frac{X\K_{, XX}+2\K_{, X}}{X\sqrt{-X} \K_{, X}}\right] \\
&& H_{2, X}= -\frac{\mu}{2} \left[  \frac{2X\K_{, \phi X}-\K_{, \phi}}{X\sqrt{-X} \K_{, X}} \right]
\eea
The contribution from the arbitrary function $h(\phi)$ is actually redundant since it enters the full Lagrangian as a total derivative. In any event, this new class of theories yields the following generalised Friedmann equation
\be \label{grav}
6 \kappa_GH^2-12\mu H^3+\K-2X\K_{, X}=\rho
\ee
and a scalar equation of motion
\be
\E_\phi \equiv -2\ddot \phi (2X\K_{, XX}+\K_{, X})-6H\dot \phi \K_{, X}+2X\K_{, \phi X}-\K_{, \phi}=0 \label{Ep}
\ee
The anomalous speed of the gravitational wave through this cosmic medium is given by
 \be
 c_T^2-1=\,{\frac {{\it \mu}}{ 2\dot \phi \left( 3\,H   {\it \mu}  -  {\it \kappa_G} \right) \K_{, X}}} \E_\phi, 
 \ee
vanishing on-shell thanks to \eqref{Ep}, as anticipated. Finally we recall the conditions for avoiding the Ostrogradski ghosts \cite{dhost3}. This places further constraints on the function, $\K$. 

\subsubsection{Adding spatial curvature} 
All of our previous analysis relied on the assumption that the cosmological metric is spatially flat. What happens when we include spatial curvature,  $k$, and proceed in a similar way? It turns out that our constraint equation \eqref{cond} receives additional terms that go as $\frac{k}{a^2} \sum_{i, j} D_{ij} H^i \dot H^j$, where the $D_{ij}(\phi, X)$ are given in terms of the (beyond) Horndeski potentials and their partial derivatives. Unless the spatial curvature is suppressed, we require all the $D_{ij}$ to vanish. However, it turns out that $D_{01}$ vanishes if and only if $G_{5, X}=0$.  As we saw earlier, this forces us back to the scenario already considered in \citeGW.  Therefore, for  the family of models given by \eqref{G2} to \eqref{F5} to be compatible with the LIGO/Virgo bounds, we require the spatial curvature of the Universe to be negligible.

\subsubsection{Constraints from decay into dark energy fluctuations}
At this stage we consider the additional constraint coming from decay of the gravitational wave into fluctuations of the scalar field \cite{decay}.  This requires the vanishing of the so-called $\frac12 \tilde m_4^2 \delta g^{00} R_{(3)} $ coupling in the effective field theory  of dark energy \cite{eft, vern}, where
\be
\tilde m_4^2=-\frac12 {\cal G}_T\alpha_T- X^2 F_4+3HX^2 \dot \phi F_5
\ee
For the class of theories given by equations \eqref{G2} to \eqref{F5}, ${\cal G}_T\alpha_T$ vanishes by the scalar equation of motion.  For $\tilde m_4^2$ to vanish we also require $- X^2 F_4+3HX^2 \dot \phi F_5=0$.  The absence of $\ddot \phi$ in this latter condition means we cannot further exploit  the vanishing of the scalar equation of motion. Rather, we are forced to set $F_4=F_5=0 $ explicitly, reducing ourselves to the Horndeski limit.  This is obtained in equations \eqref{G2} to \eqref{F5}  by setting $H_1=0, H_2=3\mu W'(\phi)\sqrt{-X}$ and $\K=\Lambda-\nu e^{W(\phi)}/X$, where $\Lambda$ and $\nu$ are constants.

\section{A newly rescued theory?} 
Let us now study the dynamics of our newly ``rescued" theory. As we have shown, this falls  within the Horndeski subclass with potentials given by
 \bea
 &&G_2=-3\mu W'''X\sqrt{-X}+\Lambda-\frac{\nu e^W}{X}, \quad  G_3=-6\mu W'' \sqrt{-X} \nonumber \\
&&G_4= \kappa_G+\frac32 \mu W'\sqrt{-X}, \quad G_5=-6\frac{\mu}{\sqrt{-X}}
 \eea
and, of course,  $F_4=F_5=0$. Notice that we have a non-trivial ``$\L_5$" contribution even in the Horndeski limit.  The structure of the theory, containing non-local operators like $1/\sqrt{-X}$,  is not especially appealing.  However, similar operators   appear in the so-called cuscaton models \cite{cus} and in the extreme relativistic limit of probe branes \cite{probes1} (see also \cite{probes2}).  Alternatively, we could  imagine them arising when we integrate out light, rather than heavy,  degrees of freedom. They are also amenable to a hydrodynamical interpretation, where, for example, an operator of the form $\nabla_\nu \phi/\sqrt{-X}$ can be interpreted as a fluid velocity \cite{hyd}.  


For this Horndeski example, the field equations simplify somewhat, giving
\bea
&& 6 \kappa_GH^2-12\mu H^3+\Lambda-3\frac{\nu e^W}{X}=\rho  \label{fried} \\
&&\E_\phi \equiv \frac{3\nu e^W}{X\dot \phi} \frac{d}{dt} \ln\left (\frac{a^2 e^W}{X}\right)=0
\eea
Upon solving the scalar equation of motion, we see that the scalar contributes an effective curvature to the cosmological evolution, such that 
 \be 6 \kappa_GH^2-12\mu H^3+\Lambda-3\nu\frac{c}{a^2}=\rho
\ee
for some dimensionful integration constant $c$  that can be assumed to be as small as we like. Following the notation of \cite{kob}, tensor and scalar fluctuations are determined by the following coefficients, 
\bea
&&{\cal F}_T={\cal G}_T=2\kappa_G-6\mu H \\
&&{\cal F}_S=-2\kappa_G\frac{\dot H}{H^2}, \quad {\cal G}_S=-3 \nu\frac{ e^{W}}{H^2X} 
\eea
which are then  required to be positive for a stable background. 

The most interesting feature of this particular  dynamics, and that which really encodes our modification to General Relativity,  is the $\mu H^3$ term.  This is also present in the generic case \eqref{grav}. For this to be relevant to the late time Universe, we require $|\mu | \sim \frac{|\kappa_G|}{H_0}$, where $H_0$ is the current Hubble scale.  Indeed, this  description can {\it only} apply to the late Universe: if $\mu$ were to retain this constant value at earlier times, the $\mu H^3$ term would dominate the dynamics over the conventional $H^2$ piece, which would ultimately be incompatible with nucleosynthesis constraints. To avoid this, one ought to take the view that this behaviour only emerges at late times.  In other words, we should really think of $\mu$ as being field dependent.  The dynamics could  then be such that it starts out negligible and remains so for much of the Universe's  history, only rising to the constant value set by the scale of dark energy at late times.  

\subsubsection{Adding local inhomogeneities}

The fact that adding spatial curvature  closes the loophole might suggest that the same is true for  wave propagation through large scale  inhomogeneities. We will now check this  explicitly and demonstrate that our loophole fails to extend beyond the cosmological background. To this end, consider propagation of a gravitational wave, $h_{\mu\nu}$,  through a background metric $\bar{g}_{\mu\nu}$
which is FRW plus small, weak-field potentials  $\Phi$ and $\Psi$ representing the inhomogeneities.  The gravitational wave is  assumed to  enter in a transverse-tracefree gauge\footnote{In this gauge the gravitational wave is assumed to be purely spatial, traceless and transverse with respect to the background FRW metric: $h_{00}=h_{0i} = h^i_{\;\;i} = 0, ~\grad_j h^j_{\;\;i}=0$. To justify this assumption,  we consider the full metric as a perturbation about FRW and perform the standard decomposition with respect to the three dimensional Euclidean group \cite{stewart}.  This contains the standard tensor mode, $\tilde h_{ij}$, which is transverse and trace-free  with respect to the background FRW metric. Redefining this as  $\tilde h_{ij}=(1-2\Psi)h_{ij}$, it follows that $h_{ij}$ is also tracefree and transverse up to terms that go as $h \del \Psi$, which are neglected under our derivative dropping assumptions, at least when computing curvature.} so that the full metric is given by 
\begin{equation}
ds^2 = -(1+2\Psi) dt^2 + a^2(1-2\Phi) (\gamma_{ij} + h_{ij}) dx^i dx^j
\label{metric_Phi_h}
\end{equation}
Before proceeding further, let us establish a hierarchy of scales,
 which can tell us which terms are most  important when computing curvature.  The scalar curvature for \eqref{metric_Phi_h} is schematically of the form
\begin{equation}
R \sim \partial^2 \bar{g}  +  (1 + \#\Phi) \partial^2 h + (1 + \# \Phi)  \partial \Phi \partial h + (1 + \#\Phi) \partial^2 \Phi h
\label{R_schematic}
\end{equation}
where $\#$ are numbers of order 1. The first term $\partial^2 \bar{g} $ is the term including linear perturbations of FRW with the potentials $\Phi$ and $\Psi$.
The other terms are those expected from including the gravitational wave  $h_{\mu\nu}$ in the transverse-traceless  gauge. Now the scale associated with changes
 in $\Phi$ and $\Psi$ is the size  of the inhomogeneity,  $r$,  while for the gravitational wave  the relevant scale is its wavelength $\lambda$. 
This means that $\partial \Phi \sim \partial \Psi \sim \Phi/r$ while $\partial h \sim h /\lambda$.  Further, the  amplitude of the gravitational wave is taken to be small compared to the amplitude of  the two
potentials, so that $\epsilon_h = h/\Phi \ll 1$.  Note that for LIGO/Virgo wavelengths $\lambda \sim 1000$km and large scale inhomogeneities $r\ge 100$Mpc, we have that
\begin{equation}
\epsilon_\lambda = \lambda /r \sim 10^{-18}
\end{equation}
The amplitude of such inhomogeneities is typically $\Phi \sim 10^{-5}$, whilst that of the wave on arrival, having travelled a distance of $40$ MPc,  is $h \sim 10^{-22}$. Therefore, when the wave is a distance $d$ from the source, we have 
\be
\epsilon_h  \sim 10^{-17} \left(\frac{40 \text{MPc}}{d}\right)
\ee
and so $\epsilon_h \ll 1$ for the vast majority of the wave's trajectory, justifying the linearised approximation. For the most part, these considerations suggest  the following hierarchy of scales in  \eqref{R_schematic}: 
\begin{equation}
  \partial^2 h  \gg  \Phi  \partial^2 h  \gg  \Phi^2 \partial^2 h \gg \partial \Phi \partial h  \gg h \partial^2 \Phi 
\end{equation}
so that  the leading terms that we shall consider are $R \sim \partial^2 \bar{g}  +  (1 + \#\Phi) \partial^2 h  + \ldots$.  In other words,
within this approximation we shall  drop any non-linear potential contributions, such as  $\Phi^2 \partial^2 h$,   and any derivatives of the potentials multiplying the gravitational wave,  such as 
$\partial \Phi \partial h$ and $(\partial^2 \Phi)h $.

In General Relativity, this leads to the gravitational wave  equation of motion 
\begin{equation} \label{GRwave}
(1-2\Psi) \left(\ddot{h}^i{}_{j} + 3H \dot{h}^i{}_{j} \right) - (1+2\Phi) \frac{\grad^2}{a^2} h^i{}_{j} = 16 \pi G T^i{}_{j}
\end{equation}
meaning that our approximation is essentially that of  geometric optics and the effective metric in  \eqref{GRwave}  is the background metric $\bar{g}_{\mu\nu}$.
This is equivalent to the fact that the gravitational wave  travels on null geodesics of  $\bar{g}_{\mu\nu}$.

Now   turn to our newly ``rescued"  Horndeski theory setting $\kappa_G=1/16 \pi G$ and assuming $\dot \phi>0$. We are only interested in the pure spin-2 gravitational wave and so  perturb the scalar as  $\phi =\bar \phi(1+\varphi)$ with $\varphi \sim O(\Phi)$.  Nevertheless, the presence of the dynamical  scalar field  can introduce new terms in our propagation equation that can be dangerously large. These include  
 terms going as  $(\partial^2 \Phi) (\partial^2h)$ whose presence we might have anticipated from the effect of spatial curvature.  Indeed,  we mentioned earlier that $\alpha_T \propto \E_\phi$ can be achieved only on flat FRW backgrounds. Once curvature $k$ is allowed, $\alpha_T$ is proportional to $k$, ${\cal G}_T\alpha_T = 18 \mu^2 \dot{H} X \frac{k}{a^2} \bigl( e^{W} \nu
- 3  \mu  H X \frac{k}{a^2} \bigr)^{-1}$. Since short-wavelength gravitational waves should not be able to feel the difference between a global curvature $k$ and a local long-wavelength curvature perturbation $\grad^2 \Phi$, we might expect a perturbative analogue of $\alpha_T$, $\delta \alpha_T \propto \grad^2 \Phi$, to obstruct the loophole.

In any event, we find that the tensor mode equation is 
\begin{multline} \label{Loophole_gravitational wave New}
(1-2\Psi) \left(\ddot{h}^i{}_{j} + 3H \dot{h}^i{}_{j} \right) \\- (1+2\Phi)(1+\delta \alpha_T) \frac{\grad^2}{a^2} h^i{}_{j} + ... = \frac{16 \pi G}{ \tilde{\cal G}_T + \delta {\tilde{\cal G}}_T} T^i{}_{j}
\end{multline}
where
\begin{align}
\hspace{-0.2cm}\tilde{\cal G}_T \delta \alpha_T &\equiv  
- 4 \tilde \mu \frac{\grad^2}{a^2} \Bigl[  \Bigl(1 + \frac{3 \mu H \dot H X}{e^{W} \nu}\Bigr) \frac{\bar\phi}{ \dot \phib} \varphi+\frac{3 \mu \dot H X}{e^{W} \nu}  \Phi \Bigr] \\
 \delta {\tilde{\cal G}}_T &\equiv  6 \tilde \mu \bigl( H \Psi + \dot{\Phi} + \frac{\phib}{ \dot \phib}  \frac{ \grad^2 }{a^2} \varphi \bigr)
 \end{align}
and $\mut = 8\pi G \mu$ and $\tilde{\cal G}_T = 8 \pi G {\cal G}_T$. Let us now estimate the size of $\delta \alpha_T$. To be relevant for the late universe, where $H \sim H_0$, we assume $\mut \sim 1/H_0$. Furthermore, typically we expect, $\dot {\bar \phi} \sim H_0 \bar \phi$,  $\dot H \sim H_0^2$ and $e^W \nu/X \sim  H_0^2/8\pi G$, the latter condition following from the Friedmann equation \eqref{fried}. It immediately follows that $\delta \alpha_T \sim \grad^2 \Phi/H_0^2 \sim \Phi /(H_0 r)^2$. For large scale inhomogeneities $H_0 r \sim {\cal O}(0.1)$ and $\Phi \sim 10^{-5}$, yielding an anomalous gravitational wave propagation of one part in a thousand or so. This is completely ruled out.

The derivation of \eqref{Loophole_gravitational wave New} makes use of the scalar dynamics by direct substitution, just as was done for the background. However, we see that it does not help. We also considered additional dynamical constraints that arise if one neglects pressure perturbation and anisotropic shear as sources of the inhomogeneity. Such constraints do not alter the qualitative result.

What about the terms we ignored  in our derivation? Compared to the leading order inhomogeneous contributions, terms such as  $\partial\Phi \partial h$ are suppressed by a factor of $ \epsilon_\lambda \sim 10^{-18}$, whilst terms such as $\partial^2 \Phi h$ are further suppressed  by  $\epsilon_\lambda^2$, so these do not affect our conclusions.  There could also be non-linear terms such as $\Phi^2 \del^2 h$. In principle these could yield corrections to $\alpha_T$ of order $\Phi^2 \sim 10^{-10}$, which are also too small to affect our conclusions. The ellipses in \eqref{Loophole_gravitational wave New} stand for further relevant terms such as  $(\partial^2 \Phi) (\partial^2h)$ in which the free indices $i$ and $j$ are not both on $h$, like  $\ddot h^{ik} \nabla_j \nabla_k \varphi$. These terms introduce gravitational birefringence.

\section{Discussion}
In this paper we identified a class of scalar tensor theories for which the anomalous gravitational wave speed vanishes {\it dynamically} on cosmological backgrounds on account of the scalar equation of motion. This reveals a potential loophole, opening up the possibility of ``rescuing" certain theories that had previously been declared to be incompatible with the LIGO/Virgo bounds. Further constraints from the decay of gravitational waves into dark energy fluctuations eliminates beyond Horndeski scenarios, leaving us with a  family of theories that fall within the   Horndeski subclass, including non-trivial  ``$L_5$" interactions.  To convincingly rule out the remaining theory, we  studied  the effect of large scale inhomogeneities and demonstrated that  any anomalous propagation of the gravitational wave could not be further eliminated by  constraints arising from the inhomogeneous equations of motion.

The starting point for our analysis was the Horndeski \cite{horn, george} and  beyond Horndeski class of scalar tensor theories \cite{bh1, bh2}. We could certainly imagine extending our procedure to include extended scalar tensor  or DHOST theories \cite{marco1, dhost1,dhost2,dhost3}, multi scalar tensor theories \cite{vish,nori}, and beyond.  Indeed, theories with more than one additional field should have a much richer structure since there are more vanishing field equations to exploit.  

Our analysis explicitly spells two important lessons: the first is that it is important to use all of the  available dynamical information when establishing the viability of a theory within a given approximation; the second is the shear power of the gravitational wave observation and its ability to constrain theories at higher order in perturbation theory.

\begin{acknowledgments}
\vskip.5cm

{\bf Acknowledgments}: 
We would like to thank C. Burrage, L. Bordin, C. Charmousis, M. Crisostomi, D. Langlois, A. Lehebel, K. Noui, C. Pitrou and I. Sawicki  for useful discussions.  Some computations in this article were checked using the tensor computer algebra packages xAct \cite{xact} and xPand \cite{xpand}.  A.P is  funded by a Leverhulme Trust
  Research Project Grant. EJC, AP and  PMS are supported by  STFC  Grant No. ST/L000393/1 and ST/P000703/1. MK was 
 funded from the European Research Council under the European Union's Seventh Framework 
Programme (FP7/2007-2013) / ERC Grant Agreement n.\,617656 ``Theories and Models of the Dark Sector: Dark Matter, Dark Energy and Gravity''.
CS received funding from the European Structural and Investment Funds 
and the Czech Ministry of Education, Youth and Sports (MSMT) (Project CoGraDS - CZ.02.1.01/0.0/0.0/15003/0000437).
\end{acknowledgments}

 \appendix
 \section{Formulae for $A_{ij}, B_{ij}$} 
Here we include explicit formulae for the non-vanishing functions $A_{ij}$ and $B_{ij}$ appearing in the scalar equation of motion, $\E_\phi$. Recall, as per the discussion in footnote \ref{signchoice}, that we have traded $\dot \phi=\sqrt{-X}$.
\bea
&&A_{00}=XG_{3,\phi X}-2XG_{2, X^2}-G_{2, X}-G_{3, \phi} \\
&&A_{10}=6\sqrt{-X}\left[2X G_{4, \phi X^2}+XG_{3, X^2}+G_{3, X}+3G_{4, \phi, X}\right] \qquad\\
&& A_{20}=12 X^3 F_{4, X^2}-6X^2 G_{5, \phi X^2}+54X^2 F_{4, X}-24X^2 G_{4, X^3} \nonumber \\
&& \quad -15XG_{5, \phi X}-48XG_{4, X^2}+36 XF_4 -6G_{4,X}-3 G_{5, \phi} \quad \\ 
&& A_{30} = -2\sqrt{-X} \left[ 6 X^3 F_{5, X^2}+33X^2 F_{5, X}+2X^2 G_{5, X^3} \right.  \nonumber \\ 
&& \left.  \qquad  \qquad +7X G_{5, X^2}+30XF_5+3 G_{5,X} \right] 
\eea
\bea
&& B_{00} =G_{2, \phi}-2XG_{2, \phi X}+X G_{3, \phi^2} \\
&& B_{10} = 6 \sqrt{-X} \left[  XG_{3, \phi X}+2XG_{4, \phi^2 X} +G_{2, X}-G_{3, \phi}\right] \\
&& B_{20} = 36 X G_{4, \phi X}-3X G_{5, \phi^2}+18X^2 F_{4, \phi}+12 G_{4, \phi} \nonumber \\
&& \quad +18 X G_{3, X}-24 X^2 G_{4, \phi X^2} +12 X^3 F_{4, \phi X}-6 X^2 G_{5, \phi^2 X}  \qquad \\
&& B_{30} = 2\sqrt{-X} \left[  -6 X^3 F_{5, \phi X}-2X^2 G_{5, \phi X^2}-18 X^2 F_{4, X} \right. \nonumber \\
&& \qquad -12 X^2 F_{5, \phi} +7X G_{5, \phi X}+36 X G_{4, X^2}-36 X F_4 \nonumber \\
&& \qquad \left. +18 G_{4, X} +9 G_{5, \phi}\right] \\
&& B_{4, 0} =-6 X \left[ 6 X^2 F_{5, X} +2X G_{5, X^2}+15 XF_5+3 G_{5, X} \right] \qquad \\
&& B_{01}=6 XG_{3, X}+12 X G_{4, \phi X}+6 G_{4, \phi} \\
&& B_{11}=12 \sqrt{-X} \left[-2X^2 F_{4, X}+XG_{5, \phi X}+4X G_{4, X^2} \right. \nonumber \\ 
&& \qquad \left.-4XF_4 +2G_{4, X}+G_{5, \phi} \right] \\
&& B_{21}= -6 X \left[  6 X^2 F_{5, X}+2X G_{5, X^2}+15 XF_5 +3 G_{5, X}\right]
\eea

\end{document}